\documentclass[a4paper]{jpconf}
\usepackage{graphicx}
\usepackage{epsfig}
\usepackage{amsmath}
\newcommand{\nn}{\nonumber}
\begin{document}
\title{Bounding the greybody factors for the Reissner-Nordstr\"{o}m black holes}

\author{Tritos Ngampitipan$^{1}$ and Petarpa Boonserm$^{2}$}

\address{$^{1}$ Department of Physics, Faculty of Science, Chulalongkorn University, Bangkok 10330, Thailand}

\address{$^{2}$ Department of Mathematics and Computer Science, Faculty of Science, Chulalongkorn University, Bangkok 10330, Thailand}

\ead{tritos.ngampitipan@gmail.com}

\begin{abstract}
A black hole can emit radiation called Hawking radiation. Such radiation seen by an observer outside the black hole differs from the original radiation near the horizon of the black hole by the so-called ``greybody factor". In this paper, the bounds of the greybody factors for the Reissner-Nordstr\"{o}m black holes are obtained. These bounds can be derived by using the $2 \times 2$ transfer matrices. It is found that the charges of black holes act as good barriers.
\end{abstract}

\section{Introduction}
A black hole was believed to have been associated with the concept that anything which entered the black hole cannot escape. In 1974, Stephen Hawking, however, showed that a black hole could indeed emit radiation, which became known as the Hawking radiation \cite{Hawking}. He discovered this radiation by studying quantum field theory in a black hole background. Hawking radiation results from one particle from pair production moving out of the black hole. According to general relativity, the spacetime around a black hole behaves as gravitational potential under which particles move. Some of the radiations are reflected back into the black hole and the rest are transmitted out of the black hole. Therefore, Hawking radiation before passing the gravitational potential is different from one after passing the potential. This difference can be measured by the so-called ``greybody factor".

There has been a number of studies devoted to calculating the greybody factors \cite{Parikh, Fleming, Lange, Fernando, Kim, esc}. Moreover, one interesting technique, the $2 \times 2$ transfer matrix, which is used to derive a rigorous bound on the greybody factors, was studied in \cite{Visser, Bogoliubov, PhD thesis}. By using this method, the bounds of the greybody factors of the four-dimensional Schwarzschild black holes was obtained \cite{pet}. In this paper, we derive the bounds of the greybody factors of the four-dimensional Reissner-Nordstr\"{o}m black holes by using the $2 \times 2$ transfer matrix.

\section{Black holes}
\subsection{The Schwarzschild black holes}
The Schwarzschild metric in four dimensions is given by
\begin{equation}
ds^{2} = -f(r)dt^{2} + f^{-1}(r)dr^{2} + r^{2}d\Omega^{2},\label{S metric}
\end{equation}
where $d\Omega^{2} \equiv d\theta^{2} + \sin^{2}\theta d\phi^{2}$ and
\begin{equation}
f(r) \equiv 1 - \frac{2GM}{r}.
\end{equation}

\begin{figure}[pb]
\centerline{\psfig{file=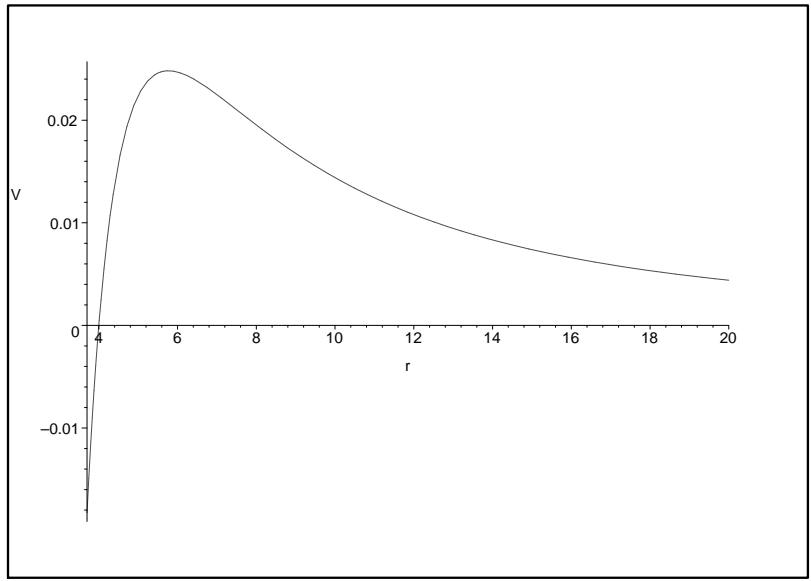,width=3in, angle = -90}}
\vspace*{1pt}
\caption{The Schwarzschild potential with $l = 1$ and $GM = 2$.}\label{S pot}
\end{figure}

\noindent The Regge-Wheeler equation is
\begin{equation}
\frac{d^{2}\psi}{dr_{*}^{2}} + \left[\omega^{2} - V(r)\right]\psi = 0,
\end{equation}
where
\begin{equation}
V(r) = f(r)\left[\frac{l(l + 1)}{r^{2}} + \frac{2GM}{r^{3}}\right],\label{VS}
\end{equation}
and the tortoise coordinate $r_{*}$ is given by
\begin{equation}
\frac{dr}{dr_{*}} = 1 - \frac{2GM}{r}.
\end{equation}
The structure of the Schwarzschild potential with $l = 1$ and $GM = 2$ is shown in Figure \ref{S pot}.

\subsection{The Reissner-Nordstr\"{o}m black holes}
The Reissner-Nordstr\"{o}m metric in four dimensions is given by
\begin{equation}
ds^{2} = -\Delta dt^{2} + \Delta^{-1}dr^{2} + r^{2}d\Omega^{2},\label{RN metric}
\end{equation}
where
\begin{equation}
\Delta \equiv 1 - \frac{2GM}{r} + \frac{GQ^{2}}{r^{2}}.\label{delta}
\end{equation}
If there are no electric charges, the Reissner-Nordstr\"{o}m metric is reduced to the Schwarzschild metric (\ref{S metric}). The only difference between the Reissner-Nordstr\"{o}m black holes and the Schwarzschild black holes is the presence of charges. The Schr\"{o}dinger like equation is
\begin{equation}
\frac{d^{2}\psi}{dr_{*}^{2}} + \left[\omega^{2} - V(r)\right]\psi = 0,
\end{equation}
where
\begin{equation}
V(r) = \frac{l(l + 1)\Delta}{r^{2}} + \frac{\Delta\partial_{r}\Delta}{r}.\label{V(r)}
\end{equation}
The tortoise coordinate $r_{*}$ is given by
\begin{eqnarray}
r_{*} = \left\{
      \begin{array}{ll}
        r + GM\ln\left|u^{2} - A^{2}\right| + \dfrac{G^{2}M^{2} + A^{2}}{2A}\ln\left|\dfrac{u - A}{u + A}\right|, & GM^{2} > Q^{2} \\
        r + GM\ln\left|u^{2} + B^{2}\right| + \dfrac{G^{2}M^{2} - B^{2}}{B}\arctan\dfrac{u}{B},                   & GM^{2} < Q^{2} \\
        r + GM\ln\left|u^{2}\right| - \dfrac{G^{2}M^{2}}{u},                                                      & GM^{2} = Q^{2} \\
      \end{array}
    \right.,
\end{eqnarray}
where
\begin{eqnarray}
u     &\equiv& r - GM\nn\\
A^{2} &\equiv& G^{2}M^{2} - GQ^{2}\\
B^{2} &\equiv& -G^{2}M^{2} + GQ^{2}.\nn
\end{eqnarray}
The black holes with $GM^{2} < Q^{2}$ are unreal and unphysical. Thus, we do not consider them. The black holes with $GM^{2} = Q^{2}$ are extreme. The extremal black holes are useful in some theories, especially in supersymmetric theories. We do not consider these black holes in this paper. We focus only on the black holes with $GM^{2} > Q^{2}$.

\begin{figure}[pb]
\centerline{\psfig{file=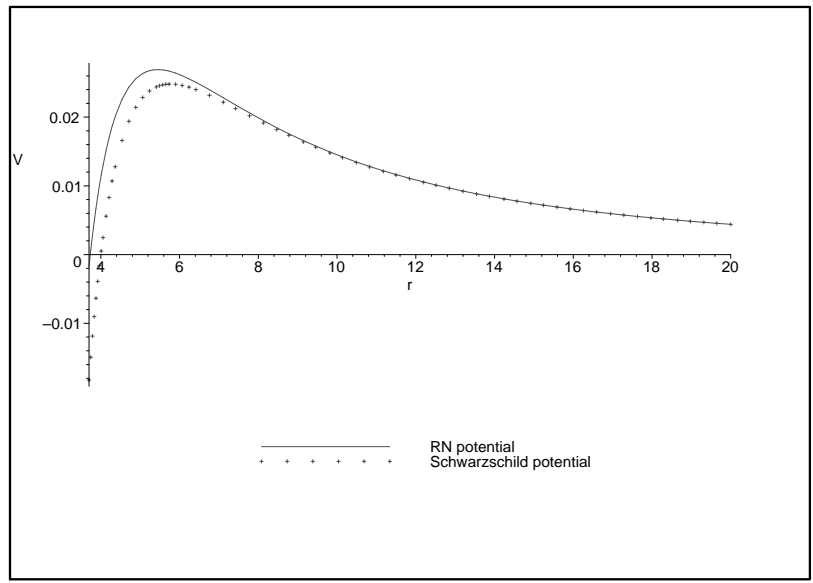,width=3in, angle = -90}}
\vspace*{1pt}
\caption{The Reissner-Nordstr\"{o}m potential with $Q = 1$, $l = 1$, and $GM = 2$ compared to the Schwarzschild potential with equal mass and angular momentum.}\label{RN pot}
\end{figure}

We can see the structure of the Reissner-Nordstr\"{o}m potential with $Q = 1$, $l = 1$, and $GM = 2$ from Figure \ref{RN pot}.

\section{The bounds of the greybody factors}
The bounds of the greybody factors using the $2 \times 2$ transfer matrix are given by \cite{Visser, Bogoliubov, PhD thesis}
\begin{equation}
T \geq \text{sech}^{2}\left(\int_{-\infty}^{\infty}\vartheta dr_{*}\right)
\end{equation}
and
\begin{equation}
R \leq \tanh^{2}\left(\int_{-\infty}^{\infty}\vartheta dr_{*}\right),
\end{equation}
where
\begin{equation}
\vartheta \equiv \frac{\sqrt{(h')^{2} + \left(\omega^{2} - V - h^{2}\right)^{2}}}{2h},\label{vartheta}
\end{equation}
for some positive function $h$. We set $h = \omega$, then
\begin{equation}
T \geq \text{sech}^{2}\left(\frac{1}{2\omega}\int_{-\infty}^{\infty}V(r)dr_{*}\right)
\end{equation}
and
\begin{equation}
R \leq \tanh^{2}\left(\frac{1}{2\omega}\int_{-\infty}^{\infty}V(r)dr_{*}\right).
\end{equation}

\subsection{The Schwarzschild black holes}
For the potential from (\ref{VS}), the bounds of the greybody factors are given by \cite{pet}
\begin{equation}
T \geq \text{sech}^{2}\left[\frac{2l(l + 1) + 1}{8GM\omega}\right]\label{TS}
\end{equation}
and
\begin{equation}
R \leq \tanh^{2}\left[\frac{2l(l + 1) + 1}{8GM\omega}\right].\label{RS}
\end{equation}

\subsection{The Reissner-Nordstr\"{o}m black holes}
For the potential from (\ref{V(r)}), the bounds of the transmission probabilities are given by
\begin{equation}
T \geq \text{sech}^{2}\left[\frac{1}{2\omega}\left\{\frac{l(l + 1)}{GM + A} + \frac{GM + 2A}{3(GM + A)^{2}}\right\}\right]\label{TRN bound}
\end{equation}
and
\begin{equation}
R \leq \tanh^{2}\left[\frac{1}{2\omega}\left\{\frac{l(l + 1)}{GM + A} + \frac{GM + 2A}{3(GM + A)^{2}}\right\}\right].
\end{equation}

\begin{figure}[pb]
\centerline{\psfig{file=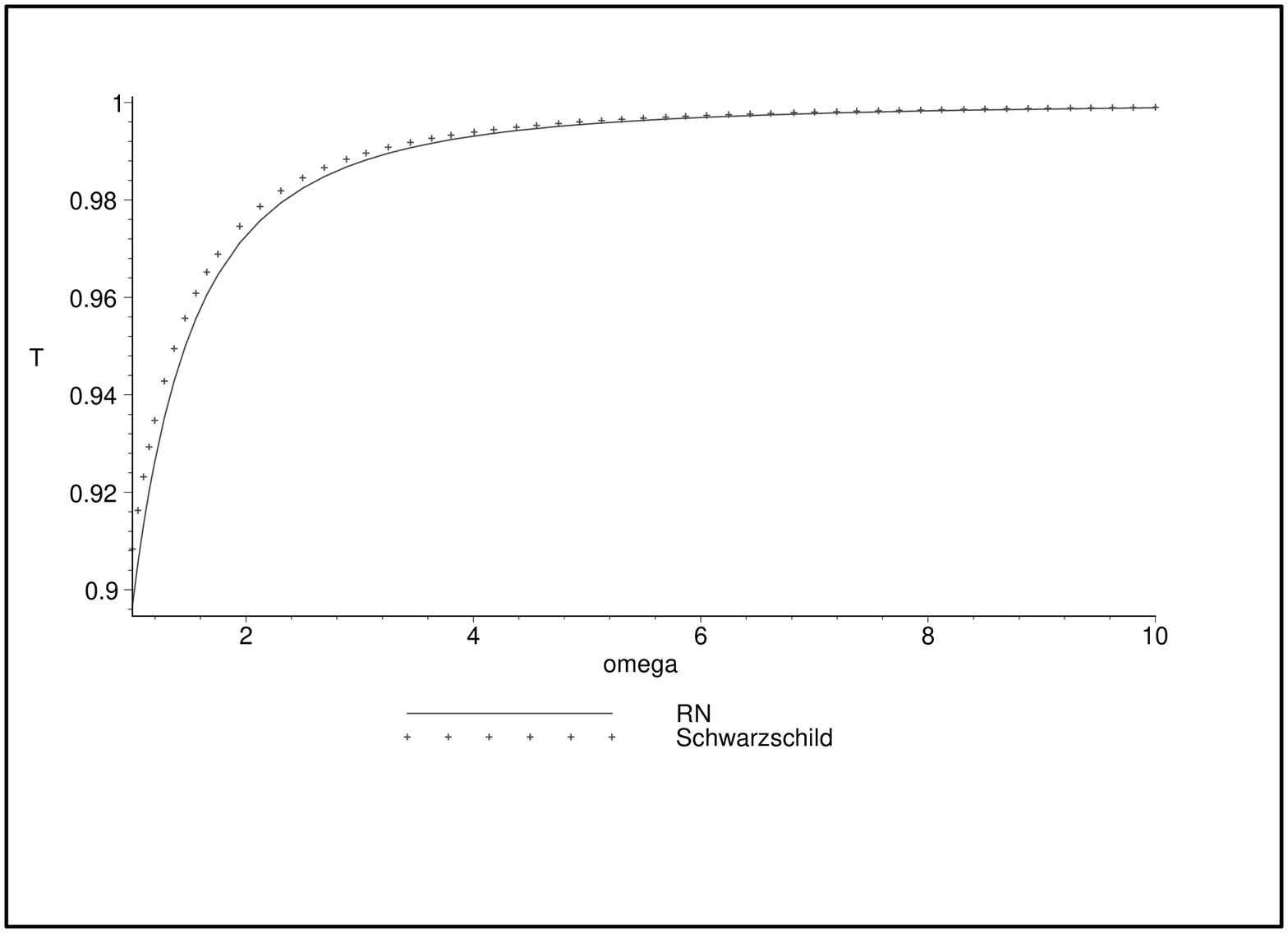,width=3in, angle = -90}}
\vspace*{1pt}
\caption{Comparison of the bounds of the transmission probabilities of the Reissner-Nordstr\"{o}m black hole and the Schwarzschild black hole with equal mass and angular momentum.}\label{TRN2}
\end{figure}

\begin{figure}[pb]
\centerline{\psfig{file=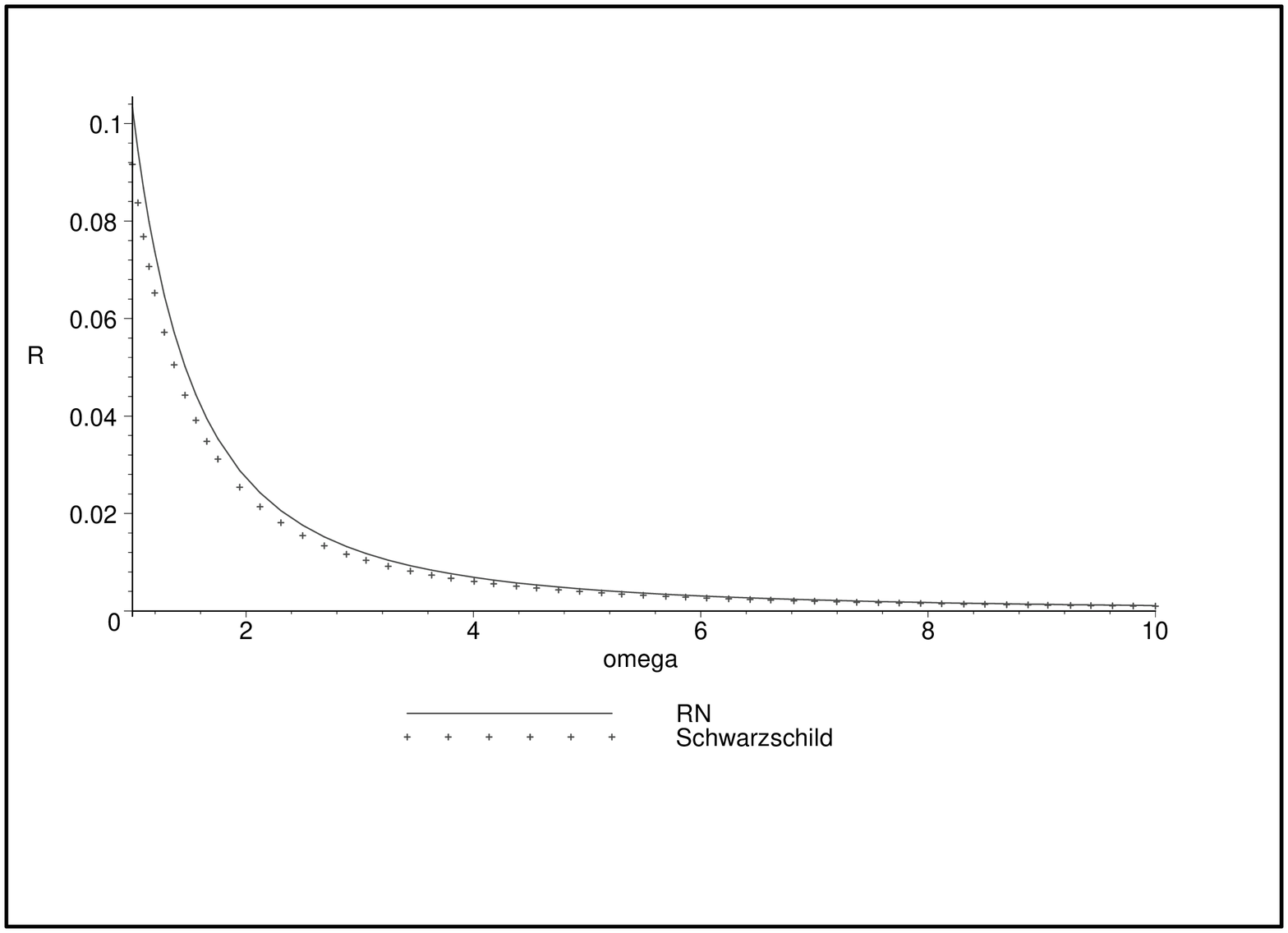,width=3in, angle = -90}}
\vspace*{1pt}
\caption{Comparison of the bound of the reflection probabilities of the Reissner-Nordstr\"{o}m black hole and the Schwarzschild black hole with equal mass and angular momentum.}\label{RRN}
\end{figure}

\noindent If the black holes have no electric charges, it is found that $A = GM$ and the above bounds are reduced to
\begin{equation}
T \geq \text{sech}^{2}\left[\frac{2l(l + 1) + 1}{8GM\omega}\right]
\end{equation}
and
\begin{equation}
R \leq \tanh^{2}\left[\frac{2l(l + 1) + 1}{8GM\omega}\right],
\end{equation}
which are exactly the bounds for the Schwarzschild black holes (\ref{TS}) and (\ref{RS}). From Figure \ref{TRN2} and \ref{RRN}, the graphs are plotted by setting $GM = 2$, $Q = 1$, and $l = 1$. The graphs show that for the Reissner-Nordstr\"{o}m black holes and the Schwarzschild black holes with equal mass and angular momentum, the greybody factor of the Reissner-Nordstr\"{o}m black holes is less than one of the Schwarzschild black holes. Since the difference between the Reissner-Nordstr\"{o}m black holes and the Schwarzschild black holes with equal mass and angular momentum is the presence of charges only, it follows that the charges of the Reissner-Nordstr\"{o}m black holes block the Hawking radiation.

\section{Conclusions}
In this paper, we obtain the bounds on the greybody factors by using the $2 \times 2$ transfer matrix for the Reissner-Nordstr\"{o}m black holes compared with ones for the Schwarzschild black holes with equal mass and angular momentum in four dimensions. It is found that the charges of the Reissner-Nordstr\"{o}m black holes behave as good barriers.

\ack
This research was supported by a grant for the professional development of new academic staff from the Ratchadapisek Somphot Fund at Chulalongkorn University, by the Thailand Toray Science Foundation (TTSF), and by the Research Strategic plan program (A1B1), Faculty of Science, Chulalongkorn University. PB was additionally supported by a scholarship from the Royal Government of Thailand. TN was additionally supported by a scholarship from the Development and Promotion of Science and Technology talent project (DPST). TN gives a special thanks to Dr. Auttakit Chatrabuti for his invaluable advice. We also thank Prof. Matt Visser for his extremely useful suggestions and comments.


\end{document}